# Coverability: Realizability Lower Bounds

Krishnendu Chatterjee[1] and Nir Piterman[2]

[1] IST Austria, Klosterneuburg, Austria
[2] University of Leicester, Leicester, UK

**Abstract.** We introduce the problem of temporal coverability for realizability and synthesis. Namely, given a language of words that must be covered by a produced system, how to automatically produce such a system. We consider the case of coverability with no further specifications, where we have to show that the nondeterminism of the produced system is sufficient to produce all the words required in the output language. We show a counting argument on a deterministic automaton representing the language to be covered that allows to produce such a system. We then turn to the case of coverability with additional specification and give a precondition for the existence of a system that produces all required words and at the same time produces only computations satisfying the additional correctness criterion. We combine our counting argument on the deterministic automaton for the language to be covered with a ranking on the deterministic Büchi automaton for the correctness criterion. One of the major issues with practical realizability is the interaction between environment assumptions and system guarantees. In many cases, synthesis produces systems that are vacuous and concentrate on forcing the environment to falsify its assumptions instead of fulfilling their guarantees. Coverability offers an alternative approach to tackle this problem.

## 1 Introduction

*Model-checking and realizability problem.* The standard *model-checking* problem given a system $S$ and specification $A$, that describes the set of *allowed* behaviors, asks whether the system satisfies the specification, i.e., the system has only allowed behaviors. In the linear-time approach, both $S$ and $A$ are considered as sets of computations and the model-checking problem translates into a language-inclusion problem $L(S) \subseteq L(A)$. That is, does the language of $A$ contain all behaviors of $S$. This is equivalent to checking whether $L(S) \cap \overline{L(A)} \neq \emptyset$, where $\overline{L(A)}$ is the set of computations *not* satisfying the specification. In case the specification $A$ is given as a deterministic automaton, its complement is easy to compute and we can use $L(\overline{A})$. If $A$ is given as a temporal-logic formula $\varphi$, then we construct directly an automaton for its complement $A_{\neg\varphi}$ [VW94]. Ultimately, model-checking is reduced to an emptiness problem of an automaton.

In the case of *synthesis and realizability* the system is expected to reply to every input of the environment by an output. The question is whether the system can do this in an online fashion (seeing one input at a time) in a way that satisfies the specification. That is, if $I$ is the set of inputs and $O$ the set of outputs, find a mapping from $I^*$ to $O$ such that the interactions resulting from interleaving an input with the mapped output satisfy the specification [PR89]. The solution calls for the construction of a two-player game modelling this interaction. If the specification $A$ is given by a deterministic automaton, we can construct from it a perfect-information turn-based two-player game $G_A$ such that the system (existential) player can *realize* some winning condition iff such a mapping exists. Choices of the system player correspond to possible outputs of the system and choice of the environment player correspond to checking all possible inputs. As the specification needs to be checked in multiple directions, deterministic (or universal, c.f., [KV05]) automata are required for the specification.

*Temporal coverability.* Model checking covers only some of the aspects of correctness (and analysis) of reactive systems. One such alternative issue is that of *coverage*. That is, both with respect to simulations and formal verification whether the tests and specifications that are written sufficient. In the case of simulation, various metrics are employed to consider whether the testing done so far exercised enough of the system. For example, one could check what parts of the code of the system was exercised or circuit covered (for hardware) (c.f., the survey [TK01]). This is similar to symbolic execution of software, an approach that is used extensively for checking security of system code [BGM13,QYPK15].

These approaches use structural properties of the system to guide testing. Alternative approaches call for trying to find types of behaviors based on semantics. For example, one can define a finite-state machine as a representation of some important features of the system and check what parts of the finite-state machine are exercised by tests (c.f., the reviews [TK01,HBB$^+$09]).

Formally stating a generalization of this approach would call for checking whether each *behavior* of a deterministic automaton describing the specification is present in the language of the system. That is, consider the dual problem of checking whether $L(A) \subseteq L(S)$. Starting from a temporal logic formula *temporal coverability* would call for testing whether each behavior allowed by a temporal logic formula is also allowed by a system. Viewing the system and the specification, again, as automata, this would call for solving the inverse language containment problem $L(A) \subseteq L(S)$. This time, as the system is nondeterministic (due to, e.g., inputs) an algorithmic solution would call for its complementation, which would clearly be intractable.

The classical approach of synthesis from linear temporal logic specifications only checks that all behaviors of the controlled system satisfy the specification. We rely on the ability of the environment to produce all inputs to ensure that the controlled system still has enough behaviors. We are not familiar with approaches that check whether the synthesized system includes a minimal set of behaviors.

*Motivating example 1.* The authors of [KPS$^+$13] applied synthesis to find a program that mimics certain aspects of a biological system. This program was then used to further study the biological system and suggest further experimental studies that could shed further light on the behavior of this system (c.f., [FPHH07,FHH11]). Specification of the system was derived from biological ex-periments, which included behaviors seen on the real live system. One aspect of the synthesis problem was to require that only experimentally observed behaviors of the system be seen in the synthesized system. However, in certain cases, the experimentally observed behavior included several options for the "same inputs". Thus, the synthesis question was generalized to also check that the produced system includes a minimum of *required* behaviors. In this case, the authors of [KPS+13] concentrated on behaviors of bounded length. The normal (bounded length) synthesis includes two quantifiers: there *exists* an implementation such that *all* its executions are correct. Here, in order to include the required behavior an additional quantifier was needed: there *exists* an implementation such that *all* its executions are correct and for *every* required behavior there *exists* a matching computation. We aim to generalize this approach beyond the case of a finite number of executions of bounded length.

*Model-checking vs temporal coverability.* As mentioned, the problems of model checking and temporal coverability are dual. The first, given system $S$ and specification $A$ calls for checking $L(S) \subseteq L(A)$. The second, calls for checking $L(A) \subseteq L(S)$. As mentioned, while the specification can be deterministic (or easily complementable through temporal logic) the system is inevitably nondeterministic. Thus, the model-checking problem is polynomial in the size of the system. However, the temporal coverability problem would be PSPACE-complete in the size of the system. This is likely to be impractical, however, the algorithmic framework within which this problem can be tackled is clear. The clarity of this algorithmic framework allows to approximate temporal coverability using testing approaches as mentioned above.

*Realizability vs temporal coverability synthesis.* When we come to extend this approach to realizability and synthesis the complexity lies elsewhere. As mentioned, for every input the system has to supply an output. If there is no restriction on the way that outputs can follow inputs the system is in some sense universal – it can produce all computations in $(I \cdot O)^\omega$. The problem, however, becomes in producing a set of required behaviors *simultaneously*. That is, is the branching of the inputs sufficient to allow writing all required outputs?

The framework of games, which is the framework providing the solution to realizability and synthesis, is no longer appropriate. We could try to seek inspiration from the game-theoretic solution to the language containment problem [WDHR06,RCDH07]. Namely, given a deterministic automaton $D$ and a nondeterministic automaton $N$ we can use a partial-information game to show that $L(D) \subseteq L(N)$. The player trying to refute the containment would choose the input word and the player trying to prove/verify the containment would choose the run of $N$ on this word. However, if the player choosing the input word knows the nondeterministic choices of the run they could cheat. Essentially, changing their mind while choosing the sequence of letters of the input word based on the nondeterministic choices made by the verifier. If the input producing refuter is blind to the nondeterministic choices made by the verifier then she cannot cheat.

This framework is, intuitively, appropriate as the system can react with every possible output to every possible input leading to nondeterminism. However, it would have to coordinate its nondeterministic choices to produce all possible behaviors. This would suggest to have an existential player choosing the mapping from inputs to outputs, another universal player choosing the input, and a third existential player choosing the word to be covered. As before, the player choosing the strategy needs to be ignorant of the actual word to be covered. Otherwise, they would be able to direct their sequence of choices to handle that specific word. In the context of full-information games the existential quantification nested within the universal quantification can be reversed resulting in just two players. However, in the context of partial-information the order of quantification cannot be reversed. This implies that we would need the context of partial-information multi-player games. Unfortunately, reasoning about partial-information multi-player games is either, in some restricted cases [KV00], of very high complexity or, in general, undecidable [PR79] (c.f., [MW03,FS05]). Hence, the framework of games does not seem to offer a solution to temporal coverability.

*Our results.* We consider the problem of temporal coverability for synthesis when the temporal property to cover is given by a deterministic finite automaton. That is, find a system $S$ of branching $I$ such that the output language of $S$ contains a specification language $L(A)$. The space of options of $S$, by definition, includes all possible sequence of outputs. So the question of whether every word in $L(A)$ can be produced by $S$ trivially holds. The complexity, intuitively, arises from the need to coordinate between very different input sequences that would need to produce related outputs in order to ensure that all possible output sequences are produced. We show that a *coverability weight* on the structure of the automaton $A$ allows us to coordinate different parts of the tree in this way.

We then turn to the problem of realizability solving both coverability and adherence to specification. That is, there is an additional correctness specification $B$ and we have to find a system $S$ of branching $I$ such that the output language of $S$ contains $L(A)$ and at the same time the input-output interactions of $S$ are contained in $L(B)$. We consider the case where $A$ is a deterministic finite automaton as before and $B$ is a deterministic Büchi automaton. Clearly, if $L(A) \nsubseteq L(B)$ the task is impossible. As $B$ is given as a deterministic automaton the check whether $L(A) \subseteq L(B)$ is simple. We extend our weighting argument and combine it with a ranking function that ensures that the correctness specification is ensured as well. In the case of coverability with correctness, we give a precondition that ensures that coverability with correctness is possible.

*Motivating example 2.* In synthesis it is very natural to partition the specification to assumptions about the behavior of the environment and guarantees of the system. This leads to a specification of the form $a \to g$. However, sometimes, the system has the ability to force the environment to falsify $a$ and void the requirement to fulfil $g$. This problem led to much research in the commuity studying synthesis suggesting various approachs to solve it. For example, in [KP10] the authors study in what cases specifications cannot lead to such situations. In [EKB15] the authors suggest how to ensure that the system cooperates with the environment. See also, [BEJK14] for a survey of different approaches to solve this problem. Coverability offers an alternative and very different approach to try and solve this problem. Consider the case of input alphabet $\{1, 2\}$ and output alphabet $\{a, b, c\}$. Suppose that the environment assumption forces the environment to input a 2 after seeing the output $a$. In addition, the environment must eventually input a 1. For an environmnent that satisfies these assumptions, the system is required to output an infinite number of $b$s. Clearly, the specification is easily realizable. However, one option for the system is to always output $a$. This implies that the environment would never be able to input 1 and the system would not have to output infinitely many $b$s. Consider the language $L = (aa\{b,c\}\{a,b,c\})^*$. We can force the system to cover this language. However, this langugae includes 6 words of length four, three with the third letter $b$ and three with the third letter $c$. It follows, that after seeing the input 2 twice the system would have to output a $b$ or a $c$ thus allowing the environmnet to input the desired 1 and forcing the system in turn to output infinitely many $b$s in return. The minimal required behavior forces the system to include enough variability in its outputs to allow the environment to fulfil its assumptions.

## 2 Background

We consider finite or infinite sequences of symbols from some finite alphabet $\Sigma$. Given a *word* $w$, an element in $\Sigma^* \cup \Sigma^\omega$, we denote by $w_i$ the $i^{th}$ letter of the word $w$, and by $w_{\geq i}$ the suffix of $w$ starting at $w_i$ hence $w = w_{\geq 0} = w_0 w_1 w_2 \ldots$. The *length* of $w$ is denoted by $|w|$ and is defined to be $\omega$ for infinite words.

**Finite Word Automata and Finite Transducers.** A *nondeterministic finite word* automaton is $N = \langle \Sigma, S, \delta, s_0, F \rangle$, where $\Sigma$ is the finite alphabet, $S$ is the finite set of states, $\delta : S \times \Sigma \to 2^S$ is the transition function, $s_0 \in S$ is the initial state, and $F$ is the acceptance set. We can run $N$ either on finite words (*nondeterministic finite word* automaton or *NFW* for short) or on infinite words (*nondeterministic Büchi word* automaton or *NBW* for short). The automaton is *deterministic* if for every state $s \in S$ and every letter $\sigma \in \Sigma$ we have $|\delta(s, \sigma)| \leq 1$. Deterministic automata are denoted DFW and DBW, for short.

A *run* of $N$ on a finite word $w = w_0, ..., w_l$ is a finite sequence of states $t_0, t_1, \ldots, t_{l+1} \in S^+$ where $t_0 = s_0$ and for every $0 \leq i \leq l$ we have $t_{i+1} \in \delta(t_i, w_i)$. A run is *accepting* if $i_m = l + 1$ and $t_m \in F$. A *run* of $N$ on an infinite word $w = w_0, w_1, ...$ is defined similarly as an infinite sequence. For an NBW, a run is *accepting* if it visits $F$ infinitely often.

A word $w$ is *accepted* by $N$ if it has an accepting run over $w$. The *language* of $N$ is the set of words accepted by $N$, denoted by $L(N)$.

A finite *transducer* $\mathcal{D}$ is $\langle \Psi, \Sigma, Q, \eta, q_0, L \rangle$, where $\Psi$ is a finite set of directions, $\Sigma$ is a finite alphabet, $Q$ is a finite set of states, $\eta : Q \times \Psi \to Q$ is a partial transition function, $q_0 \in Q$ is a start state, and $L : Q \to \Sigma$ is a labeling function. When $\eta$ is defined for every $q \in Q$ and $\psi \in \Psi$ we say that $\mathcal{D}$ is *full*. We define $\eta : \Psi^* \to Q$ in the standard way: $\eta(\varepsilon) = q_0$ and $\eta(x\psi) = \eta(\eta(x), \psi)$. Similarly, we define the extended labeling $L : \Psi^* \to \Sigma^+$ by considering $L(\varepsilon) = L(q_0)$ and $L(x\psi) = L(x) \cdot L(\eta(x\psi))$. Notice that, unless $\mathcal{D}$ is full, $\eta$ and $L$ may be partial functions. Intuitively, a transducer is a labeled finite graph with a designated start node, where the edges are labeled by $\Psi$ and the nodes are labeled by $\Sigma$. The language of a transducer $\mathcal{D}$ is the set $\{L(x) \mid x \in \Psi^*$ and $L(x)$ defined$\}$. That is, all the possible words that label paths through the graph of $\mathcal{D}$. A transducer $\mathcal{D}$ is *deterministic* if for every state $q \in Q$ and every two letters $\psi_1, \psi_2 \in \Psi$ we have if $L(\eta(q, \psi_1)) = L(\eta(q, \psi_2))$ then $\eta(q, \psi_1) = \eta(q, \psi_2)$. That is, every state $q$ and for every letter $\sigma \in \Sigma$ there is at most one successor of $q$ labeled by $\Sigma$. Given a nondeterministic transducer, it is possible to construct a deterministic transducer with the same language. This is a simple variant of the determinization of NFW [RS59].

**Trees.** Given a finite set $\Upsilon$ of directions, an $\Upsilon$-*tree* is a set $T \subseteq \Upsilon^*$ such that if $x \cdot v \in T$, where $x \in \Upsilon^*$ and $v \in \Upsilon$, then also $x \in T$. The elements of $T$ are called *nodes*, and the empty word $\varepsilon$ is the *root* of $T$. For every $v \in \Upsilon$ and $x \in T$, the node $x$ is the *parent* of $x \cdot v$. Each node $x \in T$ has a *direction* in $\Upsilon$. We assume that the root has some fixed direction $v_0 \in \Upsilon$. The direction of a node $x \cdot v$ is $v$. We denote by $dir(x)$ the direction of node $x$. An $\Upsilon$-tree $T$ is a *full infinite tree* if $T = \Upsilon^*$. Unless otherwise mentioned, we consider here full infinite trees. A *path* $\pi$ of a tree $T$ is a set $\pi \subseteq T$ such that $\varepsilon \in \pi$ and for every $x \in \pi$ there exists a unique $v \in \Upsilon$ such that $x \cdot v \in \pi$. For a node $x \in \Upsilon^*$ in the tree, we denote by $x|_i$ the prefix of $x$ up to length $i$. For example, $x|_0 = \varepsilon$ and $x|_{|x|} = x$.

Given two sets $\Upsilon$ and $\Sigma$, a $\Sigma$-*labeled* $\Upsilon$-*tree* is a pair $\langle T, \tau \rangle$ where $T$ is an $\Upsilon$-tree and $\tau : T \to \Sigma$ maps each node of $T$ to a letter in $\Sigma$. A $\Upsilon \times \Sigma$-labeled $\Upsilon$-tree $T = \langle \Upsilon^*, \tau \rangle$ is *directed* if for every $xv \in \Upsilon^+$ we have $\tau(x) \in \{v\} \times \Sigma$ and $\tau(\varepsilon) \in \{v_0\} \times \Sigma$. For a node $x \in \Upsilon^*$ the *prefix-label* of $x$ is the finite word $w$ that labels the path from the root of the tree to $x$. More formally, $|w| = |x| + 1$ and $w_i = \tau(x|_i)$. We denote the prefix-label of $x$ by $\vec{\tau}(x)$. We define the language of the tree $T$ to be $L(T) = \{\vec{\tau}(x) \mid x \in \Upsilon^*\}$.

A labeled tree is *regular* if it is the unwinding of some full transducer. Formally, a $\Sigma$-labeled $\Upsilon$-tree $\langle \Upsilon^*, \tau \rangle$ is regular if there exists a full transducer $D = \langle \Upsilon, \Sigma, Q, \eta, q_0, L \rangle$, such that for every $x \in \Upsilon^*$, we have $\tau(x) = L(x)$. The size of $\langle \Upsilon^*, \tau \rangle$, denoted $\|\tau\|$, is $|Q|$, the number of states of $\mathcal{D}$.

**Realizability.** Consider an NBW $N = \langle \Upsilon \times \Sigma, S, \delta, s_0, F \rangle$, where the alphabet has the structure $\Upsilon \times \Sigma$ for some sets of *inputs* and *outputs* $\Upsilon$ and $\Sigma$, respectively. The language of $N$ is *realizable* if there exists a full directed $\Upsilon \times \Sigma$-labeled $\Upsilon$-tree $T$ such that $L(T) \subseteq L(N)$. We say that $T$ *realizes* $N$. Notice that this implies that the initial state of $N$ reads the label of the root of the tree. As the direction of the root is $v_0$ this implies that for some $\sigma \in \Sigma$ we have $\delta(s_0, (v_0, \sigma)) \neq \emptyset$.

**Theorem 1.** [PR89] *The language of $N$ is realizable iff there exists a finite transducer $D_N$ over directions $\Upsilon$ and alphabet $\Upsilon \times \Sigma$ such that for every $q \in Q$ and every $v \in \Upsilon$ we have $L(\eta(q, v)) \in \{v\} \times \Psi$ such that $L(D_N) \subseteq L(N)$.*

## 3 Simple Covering

Solutions to realizability problems include two-player games or tree automata. The problem we are interested in here is dual. Instead (and later in addition to) of requiring that all interactions between the system and its environment produce computations that satisfy the specification, here we require that a minimal set of computations (produced by a finite transducer) are generated by the interactions.

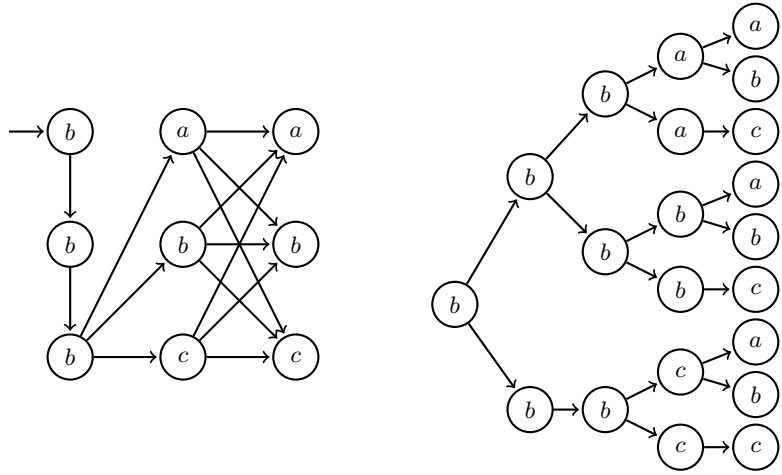

**Fig. 1.** A transducer whose language is coverable and a tree witnessing it.

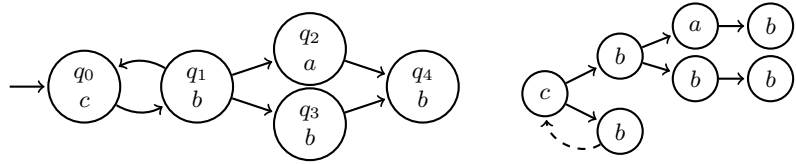

**Fig. 2.** A coverable transducer with infinite language and a tree witnessing it.

Consider an alphabet $\Sigma$, and two sets of directions $\Upsilon$ and $\Psi$. Consider a transducer $\mathcal{D} = \langle \Psi, \Sigma, Q, \eta, q_0, L \rangle$. We say that $T = \langle \Upsilon^*, \tau \rangle$ covers $L(\mathcal{D})$ if $L(T) \supseteq L(\mathcal{D})$. We note that the set $\Psi$ helps to distinguish between transitions of $\mathcal{D}$ but does not appear on $T$.

**Definition 1.** Covering. *The language of a transducer $\mathcal{D}$ is coverable with branching $\Upsilon$ if there is a labeled $\Upsilon$-tree $T$ that covers $L(\mathcal{D})$.*

Obviously, the question makes sense only in the case that $|\Upsilon| < |\Sigma|$ and $|\Upsilon| < |\Psi|$. If $|\Upsilon| \geq |\Sigma|$ then it is trivial to label an $\Upsilon$-tree by $\Sigma^*$. If $|\Upsilon| \geq |\Psi|$ we can use the regular tree induced by $\mathcal{D}$ as a subset of $\Upsilon^*$.

Consider for example the transducer in Figure 1. The language of the transducer is finite and includes only 9 words. The alphabet is $\{a, b, c\}$ and the language is $b^3\{a, b, c\}^2$, that is, all words of length 5 that start with three $b$s. We include an example of a tree of branching degree 2 (i.e., $|\Upsilon| = 2$) that covers this language. Clearly, in the case of branching degree 3 or more (i.e., $|\Upsilon| \geq 3$) the language is trivially coverable as $|\Sigma| = 3$.

We proceed with a slightly more complicated example. Consider the transducer in Figure 2. As before, the alphabet is $\{a, b, c\}$ and for now we can ignore the state names. This time, the language is infinite: $(cb)^*\{a, b\}b$. That is, start with a finite number of $cb$s and follow with an $a$ or a $b$ and a final $b$. Again, when $|\Upsilon| = 2$ there is a tree covering the language and we include such an example. As before, if $|\Upsilon| \geq 3$ then coverability is trivial.

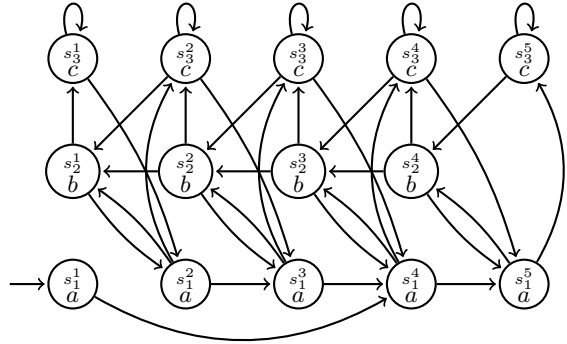

**Fig. 3.** A transducer whose language is not coverable with branching degree 2.

Finally, consider the example in Figure 3. The alphabet is $\{a, b, c\}$ as before. For a letter $\sigma \in \Sigma$ we denote $\#_\sigma(w)$ the number of occurrences of $\sigma$ in $w$. A word $u$ is a prefix of $w$ if $w = u \cdot v$ for some $v$, denoted by $u \leq w$. Then, the language of the transducer is $\{w \mid \forall u \leq w. |\#_a(u) - \#_b(u)| \leq 2 \text{ and } w_0 = a\}$. Indeed, in the states in the middle column ($s_1^3$, $s_2^3$, and $s_3^3$) and the initial state the number of $a$s and $b$s seen so far is equal. Then, an increase in the number of $a$s moves to the right. From the right-most column (states with superscript 5, where two $a$s have been seen more than $b$s, it is impossible to read further $a$s until another $b$ is read. This language is not coverable with branching degree 2. We are going to develop the tools that will enable us to prove this. But, intuitively, in a hypothetical tree covering this language there should be two nodes below the root of the tree that correspond to state $s_1^4$. Indeed, otherwise, it would be impossible to cover the three possible continuations of $aa$, namely $aaa$, $aab$, and $aac$. For exactly the same reason, two nodes in level 2 of the tree must correspond to state $s_3^4$ and two to state $s_2^3$ (the $c$-successor and the $b$-successor). As there are only four nodes in level two of the tree this leaves no free nodes to cover the runs leading to state $s_1^5$, the $a$-successor of $s_1^4$. We make this argument formal in what follows.

Consider a transducer $\mathcal{D} = \langle \Psi, \Sigma, Q, \eta, q_0, L \rangle$. We denote $(q, \psi, q') \in \eta$ for $q' = \eta(q, \psi)$. A *weight distribution* for $\mathcal{D}$ and branching $\Upsilon$ is a function $w : Q \cup (Q \times \Psi \times Q) \to \mathbb{N}$ such that the following conditions hold.

– $w(q_0) = 1$
– For every $q \in Q$ we have $|\Upsilon| \cdot w(q) \geq \sum_{(q,\psi,q') \in \eta} w(q, \psi, q')$.
  That is, the weight of $q$ can cover the weights of all outgoing transitions.
– For every $(q, \psi, q') \in \eta$ we have $w(q, \psi, q') \geq w(q')$ and $w(q, \psi, q') > 0$.
  That is, the weight of a transition can cover the target of that transition.

Intuitively, with $w(q)$ nodes in the tree we can cover the language of $\mathcal{D}_q$, namely $\mathcal{D}$ with $q$ as its initial state. The language of $\mathcal{D}_q$ is covered by allocating $w(q, \psi, q')$ of the children of the $w(q)$ nodes to successor $q'$. As $|\Upsilon| \cdot w(q) \geq \sum_{(q,\psi,q') \in \eta} w(q, \psi, q')$ we have enough children for all successors of $q$. As $w(q, \psi, q') \geq w(q')$ the children allocated for $q'$ are sufficient to cover the language of $\mathcal{D}_{q'}$.

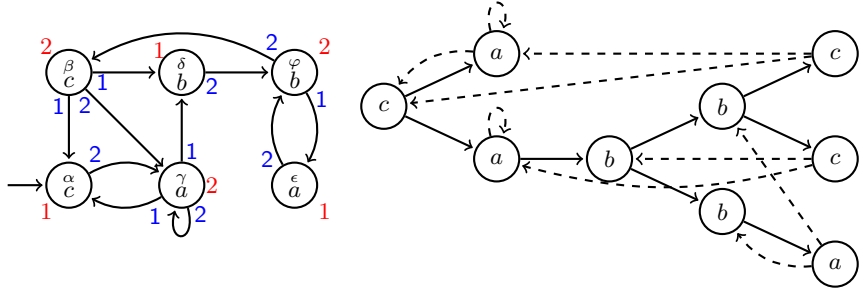

**Fig. 4.** A coverable transducer with its weight distribution (state weights in red and transition weights in blue) and the corresponding witness tree.

Consider the transducer in Figure 4. Its alphabet is $\Sigma = \{a, b, c\}$ and we consider $|\Upsilon| = 2$. The numbers written on states (red) and transitions (blue) constitute a weight distribution. Indeed, for the initial state, we have $w(\alpha) = 1$ and $2w(\alpha) \geq w(\alpha, ., \gamma)$. Also, $w(\alpha) \leq w(\beta, ., \alpha)$ and $w(\alpha) \leq w(\gamma, ., \alpha)$. For state $\gamma$, we have $w(\gamma) = 2$, which is also the weight of the three incoming transitions $(\alpha, ., \gamma)$, $(\beta, ., \gamma)$, and $(\gamma, ., \gamma)$. Also $2w(\gamma) \geq w(\gamma, ., \alpha) + w(\gamma, ., \gamma) + w(\gamma, ., \delta)$. The conditions for the other states and transitions can be verified as well.

We also go back to the transducer in Figure 3 and show that it is not coverable with branching degree 2. As $w(s_1^1) = 1$ the maximal weight for the transition to $s_1^4$ is 2 implying that $w(s_1^4) \geq 2$. Suppose that $w(s_1^4) = 2$. Then $w(s_1^4, ., s_3^4)$ must be at least 2 as well as the two states have the same language after the initial symbol. It follows that both $w(s_1^4, ., s_1^5)$ and $w(s_1^4, ., s_2^3)$ are 1. As $w(s_1^4, ., s_2^3) = 1$ it must be that $w(s_2^3) \leq 1$, which contradicts $2w(s_2^3) \geq 3$. Suppose that $w(s_1^4) = n > 2$. As before $w(s_1^4, ., s_3^4) \geq n$ implying that $w(s_1^4, ., s_2^3) < n$ and $w(s_2^3) < n$. Applying the same logic to $w(s_2^3, ., s_3^3)$, we conclude that $w(s_2^3, ., s_1^4) < n$. This contradicts $w(s_2^3, ., s_1^4) \geq w(s_1^4)$. This shows that the language of this transducer cannot be covered by a tree of branching degree 2. In fact, for every $n$, the language $a^n L(\mathcal{D})$ cannot be covered by a tree of branching degree 2.

We show that the language of $\mathcal{D}$ is coverable if and only if there exists a weight distribution for $\mathcal{D}$. Intuitively, the weight tells us how many nodes in the tree should correspond to a state of the transducer (following a certain transition).

**Theorem 2.** *The language of $\mathcal{D}$ is coverable by an $\Upsilon$-tree if and only if there is a weight distribution $w$ for $\mathcal{D}$ and branching $\Upsilon$.*

*Proof.* $\Leftarrow$ We construct an $\Upsilon$-tree covering $\mathcal{D}$ from a weight distribution.

Consider the tree $\Upsilon^*$. Based on the weight distribution $w$, we construct a $Q^+$-labeled $\Upsilon$-tree $T = \langle \Upsilon^*, \pi \rangle$ by induction, where the labeling may be partial. The labeling maintains the following:

- for every $x \in \Upsilon^*$ we have $|x| + 1 = |\pi(x)|$.

- for every run $r \in Q^+$ of $\mathcal{D}$ ending in state $q$, we maintain the invariant that there are at least $w(q)$ nodes $\{x_1, \ldots, x_{w(q)}\}$ in $\varUpsilon^*$ such that for every $i$ we have $\pi(x_i) = r$.
- for every $xv \in \varUpsilon^+$ if $\pi(xv) = rqq'$ then $\pi(x) = r \cdot q$ for some $q'$ such that $(q, \psi, q') \in \eta$.

  That is, the nodes in the tree are labeled by runs of $\mathcal{D}$ and children of a node are labeled by runs that extend the run labeling their parent.

We set $\pi(\varepsilon) = q_0$. Clearly, $|\varepsilon| + 1 = |q_0|$. By definition of weight distribution $w(q_0) = 1$ and hence one node $\varepsilon$ labeled by the run $q_0$ is sufficient.

Consider a run $r$ ending in state $q$ and let $X = \{x_1, \ldots, x_{w(q)}\} \subset \varUpsilon^*$ such that for every $i$ we have $\pi(x_i) = r$. By definition of the weight distribution we have $|\varUpsilon| \cdot w(q) \geq \sum_{(q,\psi,q') \in \eta} w(q, \psi, q')$ and for every $(q, \psi, q') \in \eta$ we have $w(q, \psi, q') \geq w(q')$. Let $Y$ denote the set of nodes $\{x \cdot v \mid x \in X \text{ and } v \in \varUpsilon\}$. It follows that $|Y| = |\varUpsilon| \times w(q)$. Hence, we can partition $Y$ such that for every $(q, \psi, q') \in \eta$ there is a set $Y_{(q,\psi,q')}$ of size $w(q, \psi, q')$ such that for every two $(q, \psi, q')$ and $(q, \psi', q'')$ such that $\psi \neq \psi'$ we have $Y_{(q,\psi,q')} \cap Y_{(q,\psi',q'')} = \emptyset$. It follows that it is safe to label the nodes $xv \in Y_{(q,\psi,q')}$ by the run $\pi(x) \cdot q'$. That is, for $xv \in Y_{(q,\psi,q')}$ we set $\pi(xv) = \pi(x) \cdot q'$.

As $w(q, \psi, q') \geq w(q')$ there are at least $w(q')$ nodes labeled by $r \cdot q'$. As $|x| + 1 = |r|$ then $|xv| + 1 = |rq'|$. Finally, $\pi(xv)$ extends $\pi(x)$ by $q'$.

This completes the construction of $\pi$ by induction. We now use $\pi$ to show a labeling $\tau$ for $\varUpsilon^*$ that covers the language of $\mathcal{D}$. By the invariant for labeling, for every run $r$ of $\mathcal{D}$ there is (at least) a node $x \in \varUpsilon$ such that $\pi(x) = r$. We set $\tau(x) = L(q)$, where $q$ is the last state of $\pi(x)$. As required, $L(\mathcal{D}) \subseteq L(T)$.

$\Rightarrow$ Consider a tree $T = \langle \varUpsilon^*, \tau \rangle$ such that $L(\mathcal{D}) \subseteq L(T)$. As $\mathcal{D}$ is deterministic, we can create a mapping $\pi : \varUpsilon^* \to Q^+$ mapping every node $x$ of $\varUpsilon^*$ to the run of $\mathcal{D}$ corresponding to $\vec{\tau}(x)$.

Let $R$ denote the set of runs of $\mathcal{D}$. For every state $q \in Q$, let $w(q)$ denote $\min_{rq \in R} |\{x \in \varUpsilon^* \mid \pi(x) = rq\}|$. That is, of all the runs ending in $q$, all of them appearing in the tree $T$, consider the one that has the minimal number of copies in $T$. For a state $q$, let $X_q = \{x_1, \ldots, x_{w(q)}\}$ denote the set of nodes of $\varUpsilon^*$ such that there is a run $rq$ of $\mathcal{D}$ and for every $i$ we have $\pi(x_i) = rq$, where $rq$ gives rise to the minimal such set. Consider a transition $(q, \psi, q')$ of $\mathcal{D}$. Clearly, all the runs $rqq'$ have to appear in $T$. Let

$$Y_{(q,\psi,q')} = \{x \cdot v \mid x \in X_q, \ v \in \varUpsilon, \text{ and } \pi(x \cdot v) = rqq'\}.$$

Clearly, as $X_q$ includes all nodes such that $r(x) = rq$ we have $Y_{(q,\psi,q')}$ includes all nodes such that $r(x) = rqq'$. We set $w(q, \psi, q') = |Y_{(q,\psi,q')}|$.

By definition $q_0$ must label the root of the tree $\varepsilon$. Hence, $w(q_0) = 1$. Furthermore, it must be the case that $\sum_{(q,\psi,q') \in \eta} w(q, \psi, q') \leq |\varUpsilon| \cdot w(q)$. Indeed, for all successors $q'$ of $(q, \psi, q')$ the weight of $(q, \psi, q')$ is induced by the same set of nodes $X_q$ in the tree and $X_q$ has at most $|X_q| \cdot |\varUpsilon|$ children in $T$. Finally, $w(q, \psi, q') \geq w(q')$ as $w(q, \psi, q')$ is chosen according to the number of children of $X_q$ that are labeled by $q'$. As $X_{q'}$ is chosen as the minimal set, it follows that $w(q, \psi, q') \geq w(q')$.

**Corollary 1.** *In a weight distribution for $\mathcal{D}$ and branching $\Upsilon$ all weights are bounded by $|\Upsilon|^{|Q|}$. It follows that the decision of whether the language of a transducer $\mathcal{D}$ is coverable is in NP.*

The proof of this corollary is included in Appendix.

We note that the tree is not restricted "from above". That is, the words that are possible to write on branches of the tree are all possible words over $\Psi^\omega$. Thus, the issue of whether the language of $\mathcal{D}$ is contained in a nondeterministic language is bypassed.

## 4 Coverability with Büchi Realizability

We extend the problem of coverability by considering simultaneously that all the paths in the tree must be labeled by words accepted by a DBW. We give a precondition for coverability that is not complete.

Consider a transducer $\mathcal{D} = \langle \Psi, \Sigma, Q, \eta, q_0, L \rangle$ and a DBW $B = \langle \Upsilon \times \Sigma, S, \delta, s_0, F \rangle$. As usual, in order to realize the DBW for every $v \in \Upsilon$ we have to choose a $\sigma \in \Sigma$ such that the infinite word resulting from such repeated interactions is in the language of $B$. We note that initially one has to choose a $\sigma \in \Sigma$ only for the initial direction $v_0$. For the transducer, $\mathcal{D}$, we choose to ignore the input $\Upsilon$. The reasoning is that the tree must include all possible options for $\Upsilon$ and there is no choice about their inclusion. In order to define covering in a way that does take the input into account we would have to allow the transducer to "choose" between different options for inputs and enforce a very unusual definition of what is the language of the transducer. The techniques we establish will form foundations for such definitions of covering as well.

**Definition 2.** Covering with Büchi Realizability. *The language of a transducer $\mathcal{D}$ is coverable while realizing DBW $B$ with branching $\Upsilon$ if there is a full directed $\Upsilon \times \Sigma$-labeled $\Upsilon$-tree that covers $\mathcal{D}$ and realizes $B$.*

As before, the directions $\Psi$ read by $\mathcal{D}$ are used to distinguish between different transitions and do not appear on the tree.

We note that if $L(\mathcal{D})\neg \subseteq L(B)$ then covering is clearly impossible. As $B$ is deterministic, the check whether $L(\mathcal{D}) \subseteq L(B)$ is easy to perform. If the upper bound language is given as a nondeterministic automaton then the very first step in checking coverability would be to check language inclusion between $\mathcal{D}$ and $B$. Here, we assume that $L(\mathcal{D}) \subseteq L(B)$ and consider a system that can output all possible output symbols at every stage of its computation. Thus, the issue of language containment in the language of a nondeterministic system is largely bypassed.

Consider for example the transducer in Figure 2 again. Let $\Upsilon = \{\alpha, \beta\}$ be the set of input letters. We add a property that dictates that if a $b$ is combined with direction $\beta$ then only $c$ can follow it and that $c$s must follow a $b$ in direction $\alpha$ infinitely often (unless completely diverging from the language of the transducer). The safety part of this condition means that if a copy of state $q_1$ appears in direction $\beta$ it can't be followed by copies of states $q_2$ and $q_3$. This means that we want the copy of state $q_1$ in direction $\alpha$ to be followed by $q_2$ and $q_3$. However, this conflicts with the liveness property telling us that $q_1$ in direction $\alpha$ should be followed infinitely often by $q_0$. In order to accommodate this requirement we must include two copies of $q_0$ with output $c$ in level 3 of the tree both of them having input $\alpha$ followed by output $b$. This is in contrast to the previous case where one copy of $q_0$ in level 3 was sufficient. In particular, the root of the tree, which corresponds to the initial state $q_0$ of the transducer and the transition $(s_0, (., c), s_1^a)$ of the DBW is covered in a different way from the nodes $\beta\alpha$ and $\beta\beta$, which also correspond to the initial state $q_0$ of the transducer and the transition $(s_0, (., c), s_1^a)$.

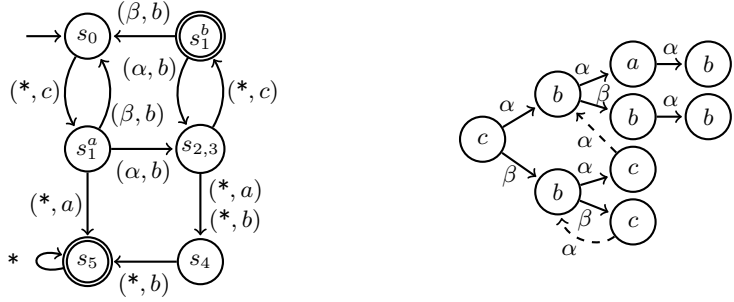

**Fig. 5.** A DBW enforcing that after a $(\beta, b)$ ($b$ in direction $\beta$) the outputs $a$ and $b$ cannot appear and that a $c$ follows a $(\alpha, b)$ (a $b$ in direction $\alpha$) infinitely often and a tree covering the transducer in Figure 2 and realizing this DBW. Directions that are missing from the tree are not restricted by the transducer and can be be easily completed in a way that realizes the DBW.

We now develop techniques similar to those developed in Section 3 for proving coverability. As notations get quite cumbersome we start with a simpler definition that shows only cover with no memory and extend it to cover with memory later.

### 4.1 Coverability without Memory

Consider a transducer $\mathcal{D} = \langle \Psi, \Sigma, Q, \eta, q_0, L \rangle$ and a DBW $B = \langle \Upsilon \times \Sigma, S, \delta, s_0, F \rangle$. We assume that $q^\perp \notin Q$ and let $Q^\perp$ denote $Q \cup \{q^\perp\}$. A combined *weight-ranking* for $B$ and $\mathcal{D}$ is $w : (Q \times S) \cup (Q \times S \times \Psi \times \Upsilon) \to \mathbb{N}$ and $d : Q^\perp \times S \to (\mathbb{N} \cup \{\infty\})$ such that the following conditions hold.

– For the weight distribution $w$ we have the following:
  • $w(q_0, \delta(s_0, (v_0, L(q_0)))) = 1$.
  • For every $q \in Q$, $s \in S$, $\psi \in \Psi$, and $v \in \Upsilon$ if $w(q, s, \psi, v) > 0$ then $\eta(q, \psi)$ is defined and $\delta(s, (v, L(\eta(q, \psi))))$ is defined.
    That is, weight can be associated only with real (combined) transitions of $\mathcal{D}$ and $B$.
  • For every $q \in Q$ and $s \in S$ we have $|\Upsilon| \cdot w(q, s) \geq \sum_{(\psi, v)} w(q, s, \psi, v)$.

That is, the weight of $w(q,s)$ is sufficient to cover the weights of all outgoing transitions.
- For every $q \in Q$, $s \in S$, and $v \in \Upsilon$ we have $\sum_{\psi \in \Psi} w(q,s,\psi,v) \leq w(q,s)$.
That is, we can use at most $w(q,s)$ successors in direction $v$.
- For every $q \in Q$, $s \in S$, $\psi \in \Psi$, and $v \in \Upsilon$ we have $w(q,s,\psi,v) \geq w(\eta(q,\psi), \delta(s,(v,L(\eta(q,\psi)))))$.
That is, the weight of a transition is sufficient to cover the target of that transition.
– For the ranking $d$ we have the following:
- $d(q_0, \delta(s_0, v_0, L(q_0))) \neq \infty$
That is, the state reached after $q_0$ reads the direction of the root and the label of $q_0$ has a finite rank.
- For every $s \notin B$ and for every $q \in Q$ and $v \in \Upsilon$ such that $d(s,q) \neq \infty$:
  * For every $\psi \in \Psi$ such that $w(q,s,\psi,v) > 0$ we have $d(q,s) > d(\eta(q,\psi), \delta(s,(v,L(\eta(q,\psi)))))$.
  * If $\sum_{\psi \in \Psi} w(q,s,\psi,v) < w(q,s)$ then there is some $\sigma \in \Sigma$ such that $d(q,s) > d(q^\perp, \delta(s,(v,\sigma)))$.
That is, if $s$ is not a Büchi state then the rank must decrease in all directions of the tree. Either by continuing to cover the transducer or by considering only realizability.
- For every $s \in B$ and for every $q \in Q$ and $v \in \Upsilon$ such that $d(s,q) \neq \infty$:
  * For every $\psi \in \Psi$ such that $w(q,s,\psi,v) > 0$ we have $d(\eta(q,\psi), \delta(s,(v,L(\eta(q,\psi))))) \neq \infty$.
  * If $\sum_{\psi \in \Psi} w(q,s,\psi,v) < w(q,s)$ then there is some $\sigma \in \Sigma$ such that $d(q^\perp, \delta(s,(v,\sigma))) \neq \infty$.
That is, if $s$ is a Büchi state then the rank must be finite in all directions of the tree. Either by continuing to cover the transducer or by considering only realizability.
- If for some $s \in S$ we have $d(q^\perp, s) \neq \infty$ then:
  * If $s \notin B$ then for every $v \in \Upsilon$ there is some $\sigma \in \Sigma$ such that $d(q^\perp, s) > d(q^\perp, \delta(s,(v,\sigma)))$.
  * If $s \in B$ then for every $v \in \Upsilon$ there is some $\sigma \in \Sigma$ such that $d(q^\perp, \delta(s,(v,\sigma))) \neq \infty$.
That is, in order to ensure realizability the ranking must be decreasing for non-Büchi states and defined for successors of Büchi states.

This definition combines the weight with a Büchi ranking, where the weight shows that the language of $\mathcal{D}$ is coverable and the ranking shows that the language of the tree realizes the language of $B$. The intuition is similar to that of the weight of the simple covering. With $w(q,s)$ nodes in the tree we can cover the language of $\mathcal{D}_q$ realizing the language of $B_s$. The ranking $d$ ensures that at the same time the language of $B$ is realizable. When the covering does not constrain the whole tree then the ranking with the "state" $q^\perp$ ensures that the language of $B$ is realizable without caring about the weights.

**Theorem 3.** *If there is a combined weight-ranking $w$ and $d$ for $B$ and $\mathcal{D}$ then the language of $\mathcal{D}$ is coverable while realizing $B$ with branching $\Upsilon$*

*Proof.* Suppose that there is a combined weight-ranking $w$ and $d$ for $B$ and $\mathcal{D}$.

Given a run $r_q = q_0, q_1, \ldots, q_n$ of $\mathcal{D}$ and a run $r_s = s_0, s_1, \ldots, s_{n+1}$ of $B$ we denote by $r_q \otimes r_s \in (Q \times S)^+$ the *joint run* $(q_0, s_1), \ldots (q_n, s_{n+1})$ (notice the omission of $s_0$). We are going to use joint runs to label the nodes of $\Upsilon^*$. We also consider extensions of runs of $\mathcal{D}$ of the form $q_0, q_1, \ldots, q_n, q^\perp, \ldots, q^\perp$, where $q_0, q_1, \ldots, q_n$ is a run of $\mathcal{D}$ and it is extended by a number of "visits" to $q^\perp$.

Consider the tree $\Upsilon^*$. Based on the weight distribution $w$ and the ranking $d$ we construct a $(Q^\perp \times S)^+$-labeled $\Upsilon$-tree $T = \langle \Upsilon^*, \pi \rangle$ by induction. The labeling maintains the following invariants:

- for every $x \in \Upsilon^*$ we have $|x| + 1 = |\pi(x)|$.
- consider a run $r_q$ of $\mathcal{D}$ ending in state $q \neq q^\perp$ such that for some run $r_s$ of $B$ ending in state $s$ there is a node $x \in \Upsilon^*$ labeled by $r_q \otimes r_s$, i.e., $\pi(x) = r_q \otimes r_s$. Then, there are at least $w(q, s)$ nodes $\{x_1, \ldots, x_{w(q,s)}\}$ in $\Upsilon^*$ such that for every $i$ there is some run $r_s^i$ of $B$ ending in $s$ such that $\pi(x_i) = r_q \otimes r_s^i$.
  That is, $r_q$ is complemented by runs $r_s$ of $B$ so that there are $w(q, s)$ copies of $r_q$ in $T$.
- for every $xv \in \Upsilon^+$ if $\pi(xv) = r(q, s)(q', s')$ then $\pi(x) = r(q, s)$ for some $q'$ and $s'$ such that either $(q, \psi, q') \in \eta$ and $(s, (v, L(q)), s') \in \delta$ or $q' = q^\perp$ and for some $\sigma \in \Sigma$ we have $(s, (v, \sigma), s') \in \delta$.
  That is, the nodes in the tree are effectively labeled by runs of $\mathcal{D}$ and $B$ and children of nodes are labeled by runs that extend the run labeling their parent.

We set $\pi(\varepsilon) = (q_0, \delta(s_0, (v_0, L(q_0))))$. Clearly, $|\varepsilon| + 1 = |(q_0, \delta(s_0, (v_0, L(q_0))))|$. By definition of weight $w(q_0, \delta(s_0, (v_0, L(q_0)))) = 1$ and hence one node $\varepsilon$ labeled by the joint run $(q_0, \delta(s_0, (v_0, L(q_0))))$ is sufficient.

Consider a run $r_q$ of $\mathcal{D}$ ending in state $q \neq q^\perp$ and a state $s$ of $B$. Let $X = \{x_1, \ldots, x_{w(q,s)}\} \subset \Upsilon^*$ be such that for every $i$ there is some run $r_s^i$ of $B$ ending in $s$ and $\pi(x_i) = r_q \otimes r_s^i$. By definition of the weight distribution we have $|\Upsilon| \cdot w(q, s) \geq \sum_{(q,s,\psi,v)} w(q, s, \psi, v)$. Let $Y$ denote the set of nodes $\{x \cdot v \mid x \in X \text{ and } v \in \Upsilon\}$. It follows that $|Y| = |\Upsilon| \times w(q, s)$.

Furthermore, for every $v \in \Upsilon$ we have (i) $\sum_{\psi \in \Psi} w(q, s, \psi, v) \leq w(q, s)$ and (ii) $w(q, s, \psi, v) \geq w(\eta(q, \psi), \delta(s, (v, L(\eta(q, \psi)))))$. It follows that we can partition $Y$ such that for every $(q, s, \psi, v)$ there is a set $Y_{(q,s,\psi,v)}$ of size $w(q, s, \psi, v)$ such that $Y_{(q,s,\psi,v)} \subseteq X \cdot \{v\}$ and for every two $(q, s, \psi, v)$ and $(q, s, \psi', v')$ such that either $\psi \neq \psi'$ or $v \neq v'$ we have $Y_{(q,s,\psi,v)} \cap Y_{(q,s,\psi',v')} = \emptyset$. Let $\overline{Y}$ denote the set of nodes in $Y$ that are not included in the subsets $Y_{(q,s,\psi,v)}$. That is,

$$\overline{Y} = Y \setminus ( \bigcup_{(q,s,\psi,v)} Y_{(q,s,\psi,v)} ).$$

It follows that it is safe to label the nodes $xv \in Y_{(q,s,\psi,v)}$ by the combined run $\pi(x) \cdot (\eta(q, \psi), \delta(s, (v, L(\eta(q, \psi)))))$. That is, for $xv \in Y_{(q,s,\psi,v)}$ we set $\pi(xv) = \pi(x) \cdot (\eta(q, \psi), \delta(s, (v, L(\eta(q, \psi)))))$.

Consider a node $xv \in \overline{Y}$. From it being in $\overline{Y}$, it follows that $\sum_{\psi \in \Psi} w(q, s, \psi, v) < w(q, s)$. Hence, for some $\sigma \in \Sigma$ we know that

$d(q^\perp, \delta(s, (v, \sigma)))$ is not $\infty$ and if needed is smaller than $d(q, s)$. We label the node $xv$ by the combined run $\pi(x) \cdot (q^\perp, \delta(v, \sigma))$. That is, for $xv \in \overline{Y}$ we set $\pi(xv) = \pi(x) \cdot (q^\perp, \delta(v, \sigma))$.

There are at least $w(\eta(q, \psi), \delta(s, (v, \eta(q, \psi))))$ nodes labeled by $r \cdot (\eta(q, \psi), \delta(s, (v, \eta(q, \psi))))$. As $|x| + 1 = |r|$ then $|xv| + 1 = |r \cdot (\eta(q, \psi), \delta(s, (v, \eta(q, \psi))))|$. Finally, $\pi(xv)$ extends $\pi(x)$ by $(\eta(q, \psi), \delta(s, (v, \eta(q, \psi))))$.

If the run $r_{q^\perp}$ of $\mathcal{D}$ ends in $q^\perp$ then by the definition of $d$ for every $v \in \Upsilon$ there is some $\sigma_v \in \Sigma$ such that $d(q^\perp, \delta(s, (v, \sigma_v)))$ is not $\infty$ and if needed is smaller than $d(q^\perp, s)$. We label the node $xv$ by $\pi(x) \cdot (q^\perp, \delta(s, (v, \sigma_v)))$. That is, $\pi(xv) = \pi(x) \cdot (q^\perp, \delta(s, (v, \sigma_v)))$.

This completes the construction of $\pi$ by induction. We now use $\pi$ to show a labeling $\tau$ for $\Upsilon^*$ that covers the language of $\mathcal{D}$ and realizes the language of $B$. By the invariant for labeling, for every run $r_1$ of $\mathcal{D}$ there is (at least) a node $x \in \Upsilon^*$ such that $\pi(x) = r_1 \otimes r_2$ for some run $r_2$ of $B$. We set $\tau(x) = L(q)$, where $q$ is the last state of $r_1$. We see that $L(\mathcal{D}) \subseteq L(T)$ as required.

Also, by the properties of $d$ it follows that every run of $B$ visits its acceptance set infinitely often. This shows that every path in the tree is accepted by $B$.

**Corollary 2.** *In a (memoryless) combined weight-ranking for $B$ and $\mathcal{D}$ and branching $\Upsilon$ all weights are bounded by $|\Upsilon|^{|Q| \times |S|}$ and all ranks by $|Q| \times |S|$. It follows that the decision of whether there exists a (memoryless) combined weight-ranking for $B$ and $\mathcal{D}$ and branching $\Upsilon$ is in NP.*

The proof of the corollary is given in Appendix.

### 4.2 Coverability with Memory

We now turn to the more complicated case where a memory is required. Notice that the example in Figure 5 requires memory. As mentioned, with the memory the notations become quite cumbersome.

Consider a transducer $\mathcal{D} = \langle \Psi, \Sigma, Q, \eta, q_0, L \rangle$ and a DBW $B = \langle \Upsilon \times \Sigma, S, \delta, s_0, F \rangle$. As before, we assume $q^\perp \notin Q$ and denote $Q^\perp = Q \cup \{q^\perp\}$.

A combined *weight-ranking* for $B$ and $\mathcal{D}$ with memory $M$ is $w : (Q \times S \times M) \cup (Q \times S \times M \times \Psi \times \Upsilon \times M) \to \mathbb{N}$ and $d : Q^\perp \times S \times M \to (\mathbb{N} \cup \{\infty\})$ such that the following conditions hold.

- For the weight distribution $w$ we have the following:
    - There is some $m_0 \in M$ such that $w(q_0, s_0, m_0) = 1$.
    - For every $q \in Q$, $s \in S$, $m, m' \in M$, $\psi \in \Psi$, and $v \in \Upsilon$ if $w(q, s, m, \psi, v, m') > 0$ then $\eta(q, \psi)$ and $\delta(s, (v, L(\eta(q, \psi))))$ are defined.
    - For every $q \in Q$, $s \in S$, and $m \in M$ we have $|\Upsilon| \cdot w(q, s, m) \geq \sum_{(\psi, v, m')} w(q, s, m, \psi, v, m')$.
    - For every $q \in Q$, $s \in S$, $m \in M$, and $v \in \Upsilon$ we have $\sum_{(\psi, m')} w(q, s, m, \psi, v, m') \leq w(q, s, m)$.
    - For every $q \in Q$, $s \in S$, $m, m' \in M$, $\psi \in \Psi$, and $v \in \Upsilon$ we have $w(q, s, m, \psi, v, m') \geq w(\eta(q, \psi), \delta(s, (v, L(\eta(q, \psi)))), m')$.

- For the ranking $d$ we have the following:
  - $d(q_0, s_0, m_0) \neq \infty$.
  - For every $s \notin B$ and for every $q \in Q$, $m \in M$, and $v \in \Upsilon$ such that $d(s, q, m) \neq \infty$:
    * For every $\psi \in \Psi$ and $m' \in M'$ such that $w(q, s, m, \psi, v, m') > 0$ we have $d(q, s, m) > d(\eta(q, \psi), \delta(s, (v, L(\eta(q, \psi)))), m')$.
    * If $\sum_{(\psi, m')} w(q, s, m, \psi, v, m') < w(q, s)$ then there is some $\sigma \in \Sigma$ and $m'' \in M$ such that $d(q, s, m) > d(q^\perp, \delta(s, (v, \sigma)), m'')$.
  - For every $s \in B$ and for every $q \in Q$, $m \in M$, and $v \in \Upsilon$ such that $d(s, q, m) \neq \infty$:
    * For every $\psi \in \Psi$ and $m' \in M'$ such that $w(q, s, m, \psi, v, m') > 0$ we have $d(\eta(q, \psi), \delta(s, (v, L(\eta(q, \psi)))), m') \neq \infty$.
    * If $\sum_{(\psi, m')} w(q, s, m, \psi, v, m') < w(q, s)$ then there is some $\sigma \in \Sigma$ such that $d(q^\perp, \delta(s, (v, \sigma)), m'') \neq \infty$.
  - If for some $s \in S$ and $m \in M$ we have $d(q^\perp, s, m) \neq \infty$ then:
    * If $s \notin B$ then for every $v \in \Upsilon$ there is some $\sigma \in \Sigma$ and $m' \in M$ such that $d(q^\perp, s, m) > d(q^\perp, \delta(s, (v, \sigma)), m')$.
    * If $s \in B$ then for every $v \in \Upsilon$ there is some $\sigma \in \Sigma$ and $m' \in M$ such that $d(q^\perp, \delta(s, (v, \sigma)), m') \neq \infty$.

This extends the previous definition with a memory value allowing to make different choices in different locations in the tree. This is required by the example in Figure 5.

**Theorem 4.** *If for some memory domain $M$ there is a combined weight-ranking $w$ and $d$ for $B$ and $\mathcal{D}$ with memory $M$ then the language of $\mathcal{D}$ is coverable while realizing $B$ with branching $\Upsilon$*

The proof, included in Appendix, adapts the ideas of the proof of Theorem 3 to include the memory.

## 5 Conclusions and Future Work

We consider the problem of realizability and synthesis with the requirement to include a minimal set of behaviors. We solve the problem in the case that there are no correctness criteria and we have to ensure that the branching of the system is sufficient to produce all required behaviors. When there are additional correctness criteria we show a precondition for the existence of a solution. We believe that these results are sufficient to merit further study of the notion of coverability in the context of realizability and synthesis.

The many future directions opened by this work include the following. Is finite-memory sufficient for coverability with realizability and is there an upper bound on the size of the required memory? Can these results be extended to more complicated specifications for realizability? How does one use coverability in practice to force more cooperative systems?

## A  Proofs of Corollary 1

We include the proof of Corollary 1:

*Proof.* The maximal cycle free run leading to a state $q \in Q$ is of length $|Q|$. It follows that in a tree covering $L(\mathcal{D})$ every state $q \in Q$ must appear before depth $|Q|$ in the tree, where the depth of $\varepsilon$ is 1. The number of nodes in depth $i$ is $|\Upsilon|^{i-1}$. The number of children of a set of nodes in depth $i$ is at most $|\Upsilon|^i$.

The bounds on $w(q)$ and $w(q, \psi, q')$ follow.

As the weights are bounded by $|\Upsilon|^{|Q|}$, the size of the entire weight is polynomial in $\Upsilon$ and $\mathcal{D}$. It is possible to check the conditions of the weight in polynomial time.

## B  Proof of Theorem 4

We include the proof of Theorem 4:

*Proof.* Suppose that there is a memory domain $M$ and a combined weight-ranking $w$ and $d$ for $B$ and $\mathcal{D}$ with memory $M$.

Given a run $r_q = q_0, q_1, \ldots, q_n$ of $\mathcal{D}$, a run $r_s = s_0, s_1, \ldots, s_{n+1}$ of $B$, and a sequence of memory values $r_m = m_0, \ldots, m_n$ we denote by $r_q \otimes r_s \in (Q \times S)^+$ the *joint run* $(q_0, s_1, m_0), \ldots (q_n, s_{n+1}, m_n)$ (notice the omission of $s_0$). We are going to use joint runs to label the nodes of $\Upsilon^*$. We also consider extensions of runs of $\mathcal{D}$ of the form $q_0, q_1, \ldots, q_n, q^\perp, \ldots, q^\perp$, where $q_0, q_1, \ldots, q_n$ is a run of $\mathcal{D}$ and it is extended by a number of "visits" to $q^\perp$.

Consider the tree $\Upsilon^*$. Based on the weight distribution $w$ and the ranking $d$ we construct a $(Q^\perp \times S \times M)^+$-labeled $\Upsilon$-tree $T = \langle \Upsilon^*, \pi \rangle$ by induction. The labeling maintains the following invariants:

- for every $x \in \Upsilon^*$ we have $|x| + 1 = |\tau(x)|$.

- consider a run $r_q$ of $\mathcal{D}$ ending in state $q \neq q^\perp$ such that for some run $r_s$ of $B$ ending in state $s$ and a sequence of memory values $r_m$ ending in memory value $m$ there is a node $x \in \Upsilon^*$ labeled by $r_q \times r_s \times r_m$. Then, there are at least $w(q,s,m)$ nodes $\{x_1, \ldots, x_{w(q,s,m)}\}$ in $\Upsilon^*$ such that for every $i$ there is a run $r_s^i$ of $B$ ending in $s$ and a sequence of memory values $r_m^i$ ending in $m$ such that $\pi(x_i) = r_q \otimes r_s^i \otimes r_m^i$.
  That is, $r_q$ is complemented by runs $r_s$ of $B$ and memory values $r_m$ so that there are $w(q,s,m)$ copies of $r_q$ in $T$.
- for every $xv \in \Upsilon^*$ if $\pi(xv) = r(q,s,m)(q',s',m')$ then $\pi(x) = r(q,s,m)$ for some $q'$, $s'$, and $m'$ such that either $(q, \psi, q') \in \eta$ and $(s, (v, L(q)), s') \in \delta$ or $q' = q^\perp$ and for some $\sigma \in \Sigma$ we have $(s, (v, \sigma), s') \in \delta$.
  That is, the nodes in the tree are effectively labeled by runs of $\mathcal{D}$ and $B$ and children of nodes are labeled by runs that extend the run labeling their parent.

We set $\pi(\varepsilon) = (q_0, \delta(s_0, (v_0, L(q_0))), m_0)$. Clearly, $|\varepsilon| + 1 = |(q_0, \delta(s_0, (v_0, L(q_0))), m_0)|$. By definition of weight there is some $m_0 \in M$ such that $w(q_0, \delta(s_0, (v_0, L(q_0))), m_0) = 1$ and hence one node $\varepsilon$ labeled by the joint run $(q_0, \delta(s_0, (v_0, L(q_0))), m_0)$ is sufficient.

Consider a run $r_q$ of $\mathcal{D}$ ending in state $q \neq q^\perp$, a state $s$ of $B$ and a memory value $m \in M$. Let $X = \{x_1, \ldots, x_{w(q,s,m)}\} \subset \Upsilon^*$ such that for every $i$ there is some run $r_s^i$ of $B$ ending in $s$ and a sequence $r_m^i$ of memory values ending in $m$ and $\pi(x_i) = r_q \otimes r_s^i \otimes r_m^i$. By definition of the weight distribution we have $|\Upsilon| \cdot w(q,s,m) \geq \sum_{(q,s,\psi,v,m')} w(q,s,\psi,v,m')$. Let $Y$ denote the set of nodes $\{x \cdot v \mid x \in X \text{ and } v \in \Upsilon\}$. It follows that $|Y| = |\Upsilon| \times w(q,s,m)$.

Furthermore, for every $v \in \Upsilon$ we have (i) $\sum_{(\psi,m')} w(q,s,m,\psi,v,m') \leq w(q,s,m)$ and (ii) $w(q,s,m,\psi,v,m') \geq w(\eta(q,\psi), \delta(s,(v, L(\eta(q,\psi)))), m')$. It follows that we can partition $Y$ such that for every $(q,s,m,\psi,v,m')$ there is a set $Y_{(q,s,m,\psi,v,m')}$ of size $w(q,s,m,\psi,v,m')$ such that $Y_{(q,s,m,\psi,v,m')} \subseteq X \cdot \{v\}$ and for every two $(q,s,m,\psi,v,m')$ and $(q,s,m,\psi',v',m'')$ such that either $\psi \neq \psi'$, $v \neq v'$, or $m' \neq m''$ we have $Y_{(q,s,m,\psi,v,m')} \cap Y_{(q,s,m,\psi',v',m'')} = \emptyset$. Let $\overline{Y}$ denote the set of nodes in $Y$ that are not included in the subsets $Y_{(q,s,m,\psi,v,m')}$. That is,

$$\overline{Y} = Y \setminus (\bigcup_{(q,s,m,\psi,v,m')} Y_{(q,s,m,\psi,v,m')}).$$

It follows that it is safe to label the nodes $xv \in Y_{(q,s,m,\psi,v,m')}$ by the combined run $\pi(x) \cdot (\eta(q,\psi), \delta(s, (v, L(\eta(q,\psi)))), m')$. That is, for $xv \in Y_{(q,s,m,\psi,v,m')}$ we set $\pi(xv) = \pi(x) \cdot (\eta(q,\psi), \delta(s, (v, L(\eta(q,\psi)))), m')$.

Consider a node $xv \in \overline{Y}$. From it being in $\overline{Y}$, it follows that $\sum_{(\psi,m')} w(q,s,m,\psi,v,m') < w(q,s,m)$. Hence, for some $\sigma \in \Sigma$ and some $m' \in M$ we know that $d(q^\perp, \delta(s,(v,\sigma)), m)$ is not $\infty$ and if needed is smaller than $d(q,s,m)$. We label the node $xv$ by the combined run $\pi(x) \cdot (q^\perp, \delta(v,\sigma), m')$. That is, for $xv \in \overline{Y}$ we set $\pi(xv) = \pi(x) \cdot (q^\perp, \delta(v,\sigma), m')$.

There are at least $w(\eta(q,\psi), \delta(s,(v, \eta(q,\psi))), m')$ nodes labeled by $r \cdot (\eta(q,\psi), \delta(s,(v, \eta(q,\psi))), m')$. As $|x| + 1 = |r|$ then $|xv| + 1 =$

$|r \cdot (\eta(q,\psi), \delta(s,(v,\eta(q,\psi))), m')|$. Finally, $\pi(xv)$ extends $\pi(x)$ by $(\eta(q,\psi), \delta(s,(v,\eta(q,\psi))), m')$.

If the run $r_{q^\perp}$ of $\mathcal{D}$ ends in $q^\perp$ then by the definition of $d$ for every $v \in \Upsilon$ there is some $\sigma_v \in \Sigma$ and some $m_v \in M$ such that $d(q^\perp, \delta(s,(v,\sigma_v)), m_v)$ is not $\infty$ and if needed is smaller than $d(q^\perp, s, m)$. We label the node $xv$ by $\pi(x) \cdot (q^\perp, \delta(s,(v,\sigma_v)), m_v)$. That is, $\pi(xv) = \pi(x) \cdot (q^\perp, \delta(s,(v,\sigma_v)), m_v)$.

This completes the construction of $\pi$ by induction. We now use $\pi$ to show a labeling $\tau$ for $\Upsilon^*$ that covers the language of $\mathcal{D}$ and realizes the language of $B$. By the invariant for labeling, for every run $r_1$ of $\mathcal{D}$ there is (at least) a node $x \in \Upsilon^*$ such that $\pi(x) = r_1 \otimes r_2 \otimes r_3$ for some run $r_2$ of $B$ and some sequence $r_3$ of memory values. We set $\tau(x) = L(q)$, where $q$ is the last state of $r_1$. We see that $L(\mathcal{D}) \subseteq L(T)$ as required.

Also, by the properties of $d$ it follows that every run of $B$ visits its acceptance set infinitely often. This shows that every path in the tree is accepted by $B$.

## C  Proof of Corollary 2

We include the proof of Corollary 2:

*Proof.* If there is a memoryless combined weight-ranking for $B$ and $\mathcal{D}$ and branchign $\Upsilon$ then every state combination $q \in Q$ and $s \in S$ that appears in the tree appears before depth $|Q| \times |S|$.

The bounds on $w(q,s)$ and $w(q,s,\psi,v)$ follow.

Consider a pair of states $(q,s)$ appearing in a location in a tree and suppose that the rank of $(q,s)$ is greater than $|Q| \times |S|$. It follows that there is a sequence of $|Q| \times |S|$ state pairs going in some direction in the tree that does not visit the Büchi set. However, this would imply that there is a loop that does not visit the Büchi set. As the ranking and weight does not include any memory the same loop will repeat forever without visitign Büchi states. It follows that all ranks are bounded by $|Q| \times |S|$.

As the weights and the ranks are bounded by $|\Upsilon|^{|Q| \times |S|}$ and the ranks are bounded by $|Q| \times |S|$ the size of the entire weight-ranking is polynomial in $\Upsilon$, $\mathcal{D}$, and $B$. It isp possible to check the conditions of the weight-ranking in polynomial time.